\documentclass[12pt]{article}
\usepackage{graphics}
\usepackage{latexsym}

\newcommand{\sect}[1]{\section{#1}\setcounter{equation}{0}}
\def\theequation{\thesection.\arabic{equation}}
\newcommand{\leaveout}[1]{}
\newcommand{\hh}{{\rm HH}}
\newcommand{\calN}{{\cal N}}
\newcommand{\calo}{{\cal O}}
\newcommand{\cals}{{\cal S}}
\newcommand{\ehat}{{\hat e}}
\newcommand{\be}{\begin{equation}}
\newcommand{\en}{\end{equation}}
\newcommand{\beq}{\begin{eqnarray}}
\newcommand{\enq}{\end{eqnarray}}

\newcommand{\fr}{\frac}

\newcommand{\eg}{{\it e.g.}}

\newcommand{\fhat}{{f_-}}

\newcommand{\fp}{{f_+}}
\newcommand{\tm}{{t_-}}
\newcommand{\tp}{{t_+}}
\newcommand{\vecm}{{\vec m}}
\newcommand{\scrim}{{\cal I}_-}
\newcommand{\scrip}{{\cal I}_+}
\newcommand{\hpl}{{\cal H}_+}
\newcommand{\hmi}{{\cal H}_-}
\newcommand{\sads}{{Schwarzschild-AdS}}
\newcommand{\roughly}[1]{\raise.3ex\hbox{$#1$\kern-.75em
\lower1ex\hbox{$\sim$}}}

\begin{document}
\bigskip
\hskip 4in\vbox{\baselineskip12pt
\hbox{hep-th/0112099}}
\bigskip\bigskip

\centerline{\Large\bf Gravitational collapse and its boundary}
\centerline{\Large\bf description
in AdS
}
\bigskip
\bigskip
\bigskip
\centerline{{\bf Steven B. Giddings}\footnote{giddings@physics.ucsb.edu} 
and {\bf Aleksey Nudelman}\footnote{anudel@physics.ucsb.edu}}
\bigskip
\bigskip
\centerline{Department of Physics, University of
California, Santa Barbara, CA\ 93106}
\bigskip

\begin{abstract}

We provide examples of gravitational collapse and black hole formation in
AdS, either from collapsing matter shells or in analogy to the
Oppenheimer-Sneider solution.  We then investigate boundary properties of
the corresponding states.  In particular, we describe the boundary
two-point function corresponding to a shell outside its horizon; if the
shell is quasistatically lowered into the horizon, the resulting state is
the Boulware state.  We also describe the more physical Hartle-Hawking
state, and discuss its connection to the quasistatic shell states and to
thermalization on the boundary.

\end{abstract}
\newpage

\baselineskip=17pt

\sect{Introduction}
The problem of quantizing gravity is perhaps most sharply focussed in the
question of what happens to quantum-mechanical information that falls into
a black hole.  Attempts to envision an answer have lead to the black hole
information paradox.\footnote{For reviews see \cite{Giddings:1994pj, 
Giddings:1995gd,Strominger:1994tn,'tHooft:ij}.}  A proposed resolution 
to the paradox that saves unitary
quantum evolution emerges from the ideas of holography\cite{tHoo,Susskind:1994vu}:
information escapes from the black hole by non-local mechanisms unique to
gravity, and can be equivalently described as being stored on the surface
of the hole.  A concrete proposal for how holography works is found in the
conjectured AdS/CFT correspondence\cite{Maldacena:1997re}, which states that string
theory in the bulk of $AdS_5\times S^5$ is equivalent to an $\calN=4$
supersymmetric gauge theory on the boundary.  

If the conjecture of \cite{Maldacena:1997re} is correct, then the gauge
theory should be capable of describing the formation and subsequent
evaporation of a black hole, and should in particular furnish a unitary
description of that process.  Understanding how this happens would
conclusively solve the black hole information paradox.

So far, however, it has been difficult to find such a concrete resolution.
The translation between bulk and boundary physics is only partially
understood; for example, it is known that correlators in the boundary
theory can be obtained from a bulk generating 
functional\cite{Gubser:1998bc,Witten:1998qj}, 
and conversely, that a bulk analog of the
S-matrix (called the ``boundary S-matrix'' in \cite{Giddings:1999qu}) can
be readily obtained from the boundary 
correlators\cite{Balasubramanian:1999ri,Giddings:1999qu}.  However, a
difficult unresolved problem has been to extract physics on scales less
than the AdS radius $R$ (or, equivalently to find the flat space S-matrix)
from this boundary S-matrix; attempts have been made in
\cite{Polchinski:1999ry,Suss}, but difficulties in obtaining flat-space physics 
have been pointed out in
\cite{Giddings:1999jq}. 

An alternate approach to trying to reconstruct  directly the flat-space
dynamics is to attempt to diagnose properties of non-trivial bulk states using
correlators on the boundary.  One might for example consider a
configuration undergoing gravitational collapse to form a black hole that
subsequently evaporates, and ask what correlation functions of boundary
operators tell us about the corresponding evolution of the gauge-theory
state on the boundary.  Better understanding of this state is particularly
relevant in confronting the information problem.  The
boundary description of a black hole formed from a pure state 
is an apparently thermal state\cite{Witten:1998qj,Witten:1998zw}, but 
in the present thinking is expected to be a fundamentally pure state.
The information hidden in this apparently thermal boundary 
state is the same as the
information hidden in the black hole in the bulk description.  Therefore
understanding how the state approaches the apparently thermal state and how
this information is encoded -- and whether it might be
accessed -- should be equivalent to understanding how information is hidden
inside a black hole, and how it might be accessed, or radiated in the
Hawking radiation.  Other discussions of the correspondence between
falling into a black hole and thermalization on the boundary have
previously been given in \cite{Banks:1998dd,Balasubramanian:1998de}.

Some preliminary aspects of the boundary description of gravitational
collapse have been investigated in \cite{Chepelev:1999zt,Giddings:1999zu} in the
case of collapse of a shell of D3 branes (corresponding to a configuration
on the Coulomb branch) to a black brane.  However, the surface of this
problem has just been scratched.  In particular, consider a family of
solutions consisting of shells of three branes with progressively smaller
radii.  In the limit when the three branes coincide, we have a macroscopic
extremal black three brane.  This limit was examined by Ross and one of the
present authors in
\cite{Giddings:1999zu}, which in particular investigated properties of two
point functions and Wilson loops in the boundary description of the
configuration.  For any finite shell radius a discrete spectrum of boundary
states was found, although it was argued that absorption effects on the
shell would broaden these states.  In the limit when the shell reaches the
horizon, these states merge into the continuum appropriate for describing
AdS in Poincar\'e coordinates.  

We'd like to understand this transition better.  It particularly becomes
puzzling when we discuss infalling observers.  For any finite radius shell,
the interior of the shell is flat space, and an infalling observer will
have to penetrate the shell in order to travel beyond.  However, the
exterior of the 
zero-radius shell is the Poincar\'e patch of AdS space, 
and we believe 
an infalling observer can sail through the horizon and explore points
beyond.  An important goal is to understand the relationship between these two
pictures.  

Of course physically speaking this story is not quite complete.  In order
to bring the individual D3 branes together we must give them a slight
velocity.  This raises the system above extremality, and a horizon will
form somewhat outside the extremal horizon; the presumed Penrose diagram
for this process was sketched in \cite{Giddings:1999zu}.  However, this
does not lessen the puzzle.  From the point of view of the gauge theory,
the coalescing D3 branes corresponds to time dependent vevs on the Coulomb
branch.  Presumably horizon formation corresponds to thermalization of this
time dependent state.  However, we should also be able to describe the
observations of an infalling observer, and after the horizon has formed
this observer will see another region of space -- behind the horizon -- 
open up for exploration.  The ``dual'' description of this physics in the
gauge theory is far from apparent.  In order to better understand the black
hole information paradox, we want to get at the root of black hole
complementarity and learn more about the map between internal degrees of
freedom of the black hole and their boundary description.

We do not yet even have the starting point of classical solutions for collapsing
D3 branes, but related 
problems exist if we return to the context of a collapsing black hole.  
For example, one can consider,
in an asymptotically AdS space,  a configuration of particles that
undergoing gravitational collapse.
A first set of questions is how to use
boundary correlators to diagnose properties of the corresponding boundary
state.  An obvious quantity to consider is the boundary stress tensor, but
in spherically symmetric collapse its expectation value is constant by
Birkhoff's theorem.  One must use finer diagnostics -- for example two- and
higher-point functions.

This paper will make some modest progress towards addressing the
question of black hole formation and the corresponding boundary
thermalization.  In particular, we will consider idealized configurations
of matter -- collapsing shells -- and discuss aspects of their
boundary description and the question of thermalization.

In outline, the next section will discuss the classical collapse, in AdS,
of a shell of dust.  (The similar problem of Oppenheimer-Snyder
collapse of a ball of dust is discussed in the appendix.)  We then turn
towards finding the boundary description of this configuration.  We wish
 to probe some of its properties 
via the bulk-boundary correspondence for a minimally coupled scalar field
moving in such a background.  Even this problem is somewhat complicated, so
we warm up with the simpler problem of computing the boundary two-point
function for a shell that we quasistatically lower to the horizon.  
Specifically, in 
section three we quantize the scalar in this background, and 
see that this has similar features to the
D3 brane shells, in particular a discrete spectrum of states.  In the limit
as the shell approaches the horizon, we show that these merge into a
continuum and we recover the Boulware
state.  In contrast to this, the correct vacuum to describe a black hole is
the Hartle-Hawking state, which we describe in section four.  The
transition between the state with a shell just outside the horizon and the
Hartle-Hawking state appears to be a difficult dynamical problem which we
still lack the tools to address.  We close with discussion of this problem
and the corresponding problem for D3 branes.  For related comments on
eternal black holes in AdS see \cite{Maldacena:2001kr}.

\sect{A collapsing shell in AdS}

We will be working in the $d+1$ dimensional spacetime of signature
$(-,+,...+)$, with action
\be
S=\fr{1}{16 \pi G} \int d^{d+1} x \sqrt{-g} (R-\Lambda) + \int d^{d+1} x
\sqrt{-g} {\cal L}_m
\en
where ${\cal L}_m$ is the matter lagrangian.
Einstein's equation are
\be
R_{\mu \nu}-{1\over2}g_{\mu \nu}R=\kappa T_{\mu \nu} - {\Lambda\over 2} 
g_{\mu \nu}\ 
\en
with $\kappa=8\pi G$.
The vacuum solution is anti-de Sitter space, which in global coordinates
has metric
\be
ds^2=-(1+r^2/b^2)dt^2+dr^2/(1+r^2/b^2)+r^2 d\Omega_{d-1}^2
\en
with 
\be
b^2=-\fr{d(d-1)}{\Lambda}\ .
\en

In order to study gravitational collapse in AdS, we consider a particularly
simple configuration:  a shell of dust, which collapses
to form a black hole.\footnote{Note that we are considering black holes in
AdS and neglect the $S^5$ factor.  Of course, a black hole smaller than the
AdS radius will be unstable\cite{Gregory:1993vy} and will develop structure
in the $S^5$ directions.  The CFT origin of the corresponding physics is
poorly understood.  We will neglect this instability, which is an added
complication, but which we believe shouldn't affect the issues of principle
which we are attempting to address.}
We can think of this as an idealization of a
collection of particles (\eg\ dilatons or massive particles) 
distributed over a thin
inward-moving shell.  Such shells have been previously considered in this
context in
\cite{Danielsson:1999zt,Danielsson:2000fa}.

The energy momentum tensor for a spherically-symmetric 
shell of pressureless dust following a radial trajectory 
$R(\tau)$ and with density
$\sigma$ is given by 
\be
T_{\mu \nu}=\sigma U_\mu U_\nu \delta (r-R(\tau))\ ,
\en
where $U^\mu$ is the velocity of the shell, which satisfies $U^\mu U_\mu = -1$.
Inside the shell the metric is that of global AdS,
\be
ds^2=-\fhat(r)dt_-^2+dr^2/\fhat(r)+r^2 d\Omega_{d-1}^2\ ,\label{gloads}
\en
where
\be
\fhat(r)= 1 +{r^2\over b^2}\ .
\en
The external metric is \sads:
\be
ds^2= -\fp(r)d\tp^2+
dr^2/\fp(r)+r^2 d\Omega_{d-1}^2\ \label{adsc}
\en
with
\be
\fp(r) = 1+{r^2\over b^2}-{m\over r^{d-2}}\ .
\en
Here the mass parameter $m$ is related to the mass by
\be
M_{ADM}= {(d-1) \Omega_{d-1} \over 16\pi G} m\ ,
\en
with $\Omega_{d-1}$ the area of the unit $d-1$ sphere, and the
Schwarzschild radius is given by
\be
m=r_H^{d-2}(1+\fr{r_H^2}{b^2})\ .
\en
The Penrose diagram for \sads\ is shown in fig.~1.
\begin{figure}
\centering
\scalebox{0.5}{\includegraphics{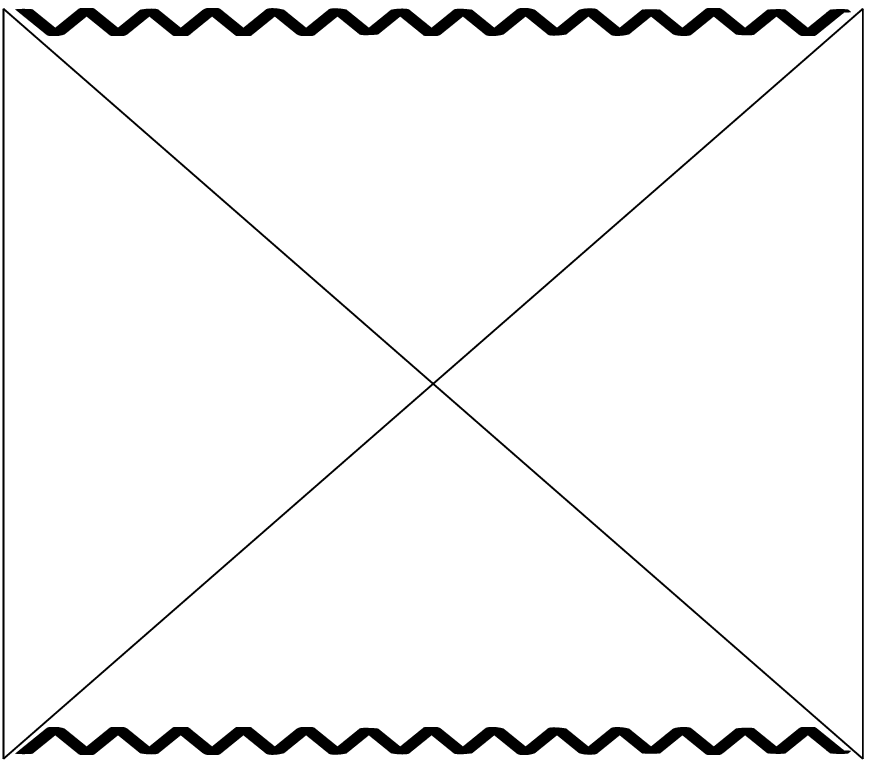}}
\caption{The Penrose diagram for \sads. }
\end{figure}

The induced metric on the surface of the shell should be the same computed
from the inside or outside metric.  In addition, if the shell is freely
falling, its motion is determined by matching the 
extrinsic curvature of the geometry across the shell.  The
first condition relates time inside and outside the shell:
\be
{d{\tm}\over d\tp} = \sqrt{\fp\over \fhat}\ .
\en 
The matching conditions for the extrinsic curvature are the Israel matching
conditions\cite{Israel:1966rt}.  These take the form
\beq
K^+_{ab}-K^-_{ab      }=-\kappa \sigma \left(U_a U_b + {g_{ab}\over 2} \right)
\label{curvmatcho}\\
(K^+_{ab}+K^-_{ab})U^a U^b=0 \label{curvmatch}\ 
\enq
where latin indices denote indices tangent to the
shell's world-volume, and $K_{ab}$ is 
the extrinsic curvature of the shell's world-volume.
These equations can be thought of as summarizing the balance of normal
forces on the shell's surface layer.
Let $n^\mu$ denote the normal vector to 
an infinitesimal element of the shell.
One can readily show that the geodesic equation of this element
implies\cite{Israel:1966rt,Kuchar} 
\be
n_\mu {DU^\mu \over d\tau}\vert_{\pm} = -K_{ab}^\pm U^a U^b\ .
\en
Combining this equation with (\ref{curvmatcho}), (\ref{curvmatch}) 
then gives the equations
\beq
n_\mu \fr{DU^\mu}{d \tau}\vert_+ +n_\mu \fr{DU^\mu}{d\tau} |_-=0 \ 
,\label{israelone}\\ 
n_\mu \fr{DU^\mu}{d\tau} |_+- n_\mu\fr{DU^\mu}{d\tau} |_-= {\kappa\sigma\over
2}\ .\label{israelmatch}
\enq
In terms of the trajectory $R(\tau)$, the velocity and normal are easily
seen to take the form
\beq
U^\mu_\pm = (\dot{t}_\pm,\dot{R},0,\ldots,0)\ ,\ 
n^\pm_\mu=(-\dot{R},\dot{t}_\pm,\ldots,0)
\enq
where
\be
\dot{t}_\pm=\sqrt{f_\pm(r)+\dot{R}^2}/f_\pm(R)\ .
\en
Given these expressions and the metric (\ref{gloads}), (\ref{adsc}), 
we may evaluate the
left hand sides of (\ref{israelone}), 
(\ref{israelmatch}), and derive a differential equation
whose first integral of the motion is
\be
1+\left (\frac{dR}{d\tau}\right )^2+\left(\fr{R}{b} \right)^2=
\left (a+\fr{m}{4aR^{d-2}} \right)^2 \\
\en
where $a$ is an integration constant.  The rest mass of the shell, if
infinitely dispersed, is 
\be
M=  {m\Omega_{d-1}(d-2)\over \kappa a} \ . 
\en
In Minkowski
space ($b\rightarrow \infty$) an expanding shell escapes to infinity if $a>1$ and
rebounds at finite radius if $a<1$.  In AdS, all expanding shells clearly
rebound at some finite radius.

The collapsing shell crosses the horizon when $f_+(R)=0$, and thus forms a
black hole, as shown in fig.~2.
\begin{figure}
\centering
\scalebox{0.5}{\includegraphics{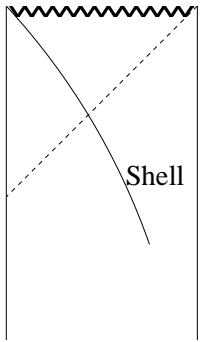}}
\caption{The Penrose diagram for collapsing shell in anti de-Sitter space. }
\end{figure}
We would like to better understand the boundary CFT
representation of this process.

The coarsest diagnostic of a configuration in AdS is its boundary stress
tensor.  While many configurations in AdS have a non-trivial boundary
stress tensor which encodes important data about the bulk state, by
Birkhoff's theorem all spherically symmetric solutions in AdS are
asymptotically of the form (\ref{adsc}) and thus have the same stress
tensor \cite{Horowitz},
\be
T_{\mu\nu} = \rho\left( t_\mu t_\nu\ + {1\over {d-1}}h_{\mu\nu} \right),
\en
where $t^\mu$ is the unit timelike Killing vector on the boundary of
\sads, $h_{\mu\nu}$ is the metric on $S^{d-1}$, and $\rho$ is the
mass density on the boundary.  The latter is related to bulk quantities by 
\be
\rho=\fr{r_H}{\kappa b^2}
\en
for $d=3$, and
\be
\rho=\fr{3}{8 \kappa b^3}(b^2+4r_H^2)
\en
for $d=4$.
Thus in particular, the boundary stress tensor does not distinguish between
a black hole and a collapsing shell.  For this reason we must seek refined
diagnostics to distinguish these configurations.  Perhaps the simplest such
example is the two-point function of fields propagating in the bulk
background, and the corresponding boundary correlator.

\sect{Quantization in the shell background}

\subsection{Classical solutions}

We next turn to the quantization of fields moving in the background of a
shell; for simplicity we consider a minimally coupled scalar field,
\be
\Box \phi=0.
\label{wave}
\en 
Also for simplicity, we consider solutions to this equation in the
background of a {\it static} shell fixed at constant $R$; this is of course
a good approximation for the slowly-moving shell, but not for example for a
shell crossing a horizon.  We can alternately consider a quasistatic
(non-geodesic) family of shells that gradually approaches the horizon;
although such a family would have to be accelerated by an external agent,
study of the corresponding bulk and boundary states is hoped to yield
further insight into gravitational collapse.

The solutions to (\ref{wave}) in the shell background are
not known, but we will infer some of their properties.  We are particularly
interested in the spectrum, and 
other properties of the state as the horizon forms.

In order to study eq.~(\ref{wave}), we will write the metric
(\ref{gloads}), (\ref{adsc})
both inside and outside the shell in the uniform form
\be
ds^2=-\tilde{f}(r)dt^2+\fr{dr^2}{f(r)}+r^2 d\Omega^2
\en
where
\beq
f&=& 1+r^2 \label{fdefl}\\
\tilde{f}&=&(1+r^2)\fr{(1-m/R^{d-2}+R^2)}{1+R^2}
\label{insidef}
\enq
for $r<R$ and
\beq
\tilde{f}=f=1-m/r^{d-2}+r^2 \label{fdefg}
\enq
for $r>R$.  Here and for the rest of the section we work in units in which the AdS
radius $b=1$.  
Notice in particular that while (\ref{insidef}) corresponds to unperturbed
AdS, the extra factor accounts for a relative redshift:  a given proper
frequency measured by an observer in the center of the AdS portion
corresponds to a redshifted frequency as seen by an asymptotic observer.

We consider a solution of (\ref{wave}) of definite frequency and angular momentum,
\be
\phi= \varphi_{\omega l\vecm}(r) e^{-i\omega t}Y_{l\vecm}\ .\label{schmode}
\en
Properties of such solutions are most easily understood by introducing
the tortoise coordinate $r_*$,
\be
dr_{*}=\fr{dr}{\sqrt{f \tilde{f}}}\ ,
\en
in terms of which the $t$, $r$ part of the metric takes conformally flat
form.  We also 
rescale the wavefunction,
\be
\varphi_{\omega l\vecm}(r) = {u_{\omega l\vecm}\over r^{(d-1)/2} }\ .
\en
In these variables the radial solution to the wave equation is found by
solving a one-dimensional potential problem, 
\be
\fr{d^2 u_{\omega l\vecm} }{dr_*^2}+(\omega^2-V_{eff})u_{\omega l\vecm}=0
\label{radeqn}
\en
with effective potential given by 
\be
V_{eff}=\fr{\tilde{f}l(l+d-2)}{r^2}+\left (\fr{d-1}{2}
\right){1\over 2r}\fr{d(f\tilde{f})}{dr}+\left(\fr{d-1}{2} \right ) \left
(\fr{d-3}{2}\right)\fr{f \tilde{f}}{r^2}\ .\label{effpot}
\en
The solutions should of course be continuous at the shell; integrating
(\ref{radeqn}) then yields the matching condition
\beq
& &\fr{d u_{\omega l\vecm}}{dr_*}|_+-\fr{du_{\omega 
l\vecm}}{dr_*}|_{-}\label{bcone}\\ 
& &=u_{\omega l\vecm}(R){d-1\over 2R}\sqrt{1-\fr{m}{R^{d-2}}+R^2}
\left(\sqrt{1-\fr{m}{R^{d-2}}+R^2}-\sqrt{1+R^2} \right ).\nonumber
\enq
This equation is easily seen to correspond to the condition that the normal
derivatives match across the shell,
\be
n^\mu \nabla_\mu \phi|^+-n^\mu \nabla_\mu\phi|^-=0\ .\label{bctwo}
\en
Note that both our effective potential (\ref{effpot}) and boundary
conditions (\ref{bcone}), (\ref{bctwo}) differ from those in
\cite{Danielsson:2000fa}.  For example, \cite{Danielsson:2000fa} replaces
the interior of the shell by the {\it Poincar\'e} patch of AdS.

The potential (\ref{effpot}) takes the form shown in fig.~3; inside the
shell it is the potential of unperturbed AdS -- with additional redshift --
and outside it is that of \sads.  The solutions will have quantized
frequencies $\omega_n$.  
\begin{figure}
\centering
\scalebox{0.7}{\includegraphics{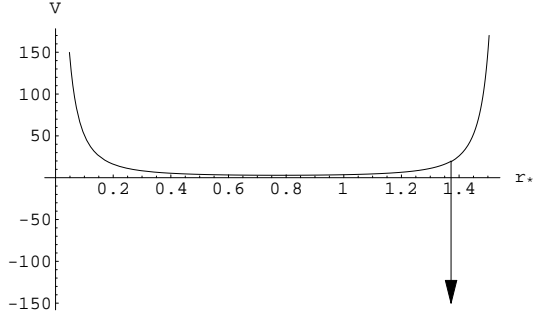}}
\caption{The effective potential for $d=4$, $l=0$, 
$r_H=2$ and $R=5$. The shell lies at $r_*\approx 1.37$ and produces a delta
function in the potential, as shown.}
\end{figure}

To see the effects of the shell on the spectrum, first consider the
situation where the shell is far from the horizon.  This case is most
easily analyzed by working in the proper time coordinate of the {\it
central} observer, which is 
\be
{\hat t}= \sqrt{ \fr{(1-m/R^{d-2}+R^2)}{1+R^2}} t\ .
\en
In this coordinate the metric takes the form
\be
ds^2=-\tilde{f}(r)d{\hat t}^2+\fr{dr^2}{f(r)}+r^2d\Omega^2\ ,
\en
where for $r<R$ 
\be 
f=\tilde{f}=1+r^2
\label{ads32}
\en
and for $r>R$
\be
f=1-{m\over r^{d-2}} + r^2\ ,\ 
\tilde{f}=(1-\fr{m}{r^{d-2}}+r^2)\fr{1+R^2}{1-\fr{m}{R^{d-2}}+R^2}\ .
\en
The wave equation then becomes
\beq
\fr{d^2 u_{n l\vecm} }{d r_*^2}+(\hat{\omega}^2-V_{eff})u_{n l\vecm}=0\ 
\enq
where $V_{eff}$ is found as in (\ref{effpot}).
This can be solved for the characteristic frequencies $\hat\omega_n$;
the 
frequencies seen by the asymptotic observer are then given by
\be
\omega_n= \sqrt{ \fr{(1-m/R^{d-2}+R^2)}{1+R^2}} \hat\omega_n\ .
\label{redshift22}
\en

The resulting effective 
potential can be written as that of AdS, plus a perturbation,
\be
V_{eff} = V_{AdS} + V_m\ .
\en
The unperturbed eigenfunctions of the AdS wave operator take the form
\be
u_{n l}(r_*)=c_{nl} \cos ^{\fr{1+d}{2}}(r_*) \sin
^{\fr{d-1+2l}{2}}(r_*) P_n^{l-1+\fr{d}{2},{d\over 2}}(\cos 2r_*)\ ,
\en 
where $c_{nl}$ are  normalization constants,
and have quantized frequencies
\be
{\hat \omega}_{nl}=d+l+2n\ .
\en

For $r<R$, the metric is precisely of AdS form, so $V_m$=0 for $r<R$.  The
perturbation for $r>R$ is bounded as
\beq
|V_m| \roughly< \fr{m}{R^{d-2}}\ .
\label{bound3}
\enq
Using  $dr_{*} \sim dr/r^2$, $u\sim \fr{1}{r^{(d+1)/2}}$ and
(\ref{bound3}), we 
obtain the shift in frequency as seen by an AdS observer:
\be
|\Delta \hat{\omega}| \roughly< \fr{m}{R^{2d}}\ .
\en
This shows that the contribution of the $r>R$ perturbation to  the
effective 
potential is subleading to the redshift contribution (\ref{redshift22}).
An additional contribution to the effective potential comes from the delta 
function at $r=R$:
\be
V_m = -\fr{(d-1)m}{4R^{d-1}}(1+R^2) \delta (r-R)\ .
\en
This gives a contribution ${\cal O}\left(\fr{m}{R^{2d}}\right)$ 
to $\Delta \hat{ \omega}$
which is 
again subleading.
Thus to leading order in $\fr{1}{R}$, we find that the frequencies seen by
the outside observer are
shifted homogeneously as
\be
\Delta\omega_{n l\vecm}=-\fr{m}{2R^{d}}\omega_{n l\vecm}\ .
\en
The net effect is a redshift of the eigenfrequencies of
the modes.  While this perturbative analysis will fail as the shell
approaches the horizon, the qualitative effect remains the same, namely
increasing redshift for decreasing radius.  When the shell reaches the
horizon, the modes of the internal AdS region are infinitely redshifted,
and the outside observer sees the emergence of a continuum.  This is clear
from the form of the effective potential; as $R$ approaches the horizon,
the shell moves to $r_* = -\infty$, and the potential
takes the form shown in fig.~4.

\begin{figure}
\centering
\scalebox{0.7}{\includegraphics{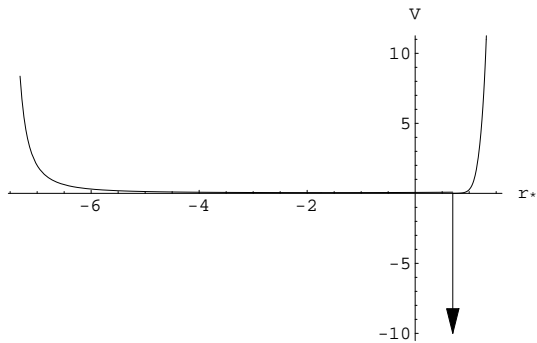}}
\caption{The effective potential for $d=4$, $l=0$,$r_h=2$ and $R=2.01$. 
The shell is close to its gravitational radius. Note that the potential in
the vicinity of the shell is very small as compared to the case of a
large-radius shell, fig.~3.  
It has also been shifted leftwards; now the shell lies
at $r_* \approx 0.7$.}
\end{figure}

\begin{figure}
\centering
\scalebox{0.7}{\includegraphics{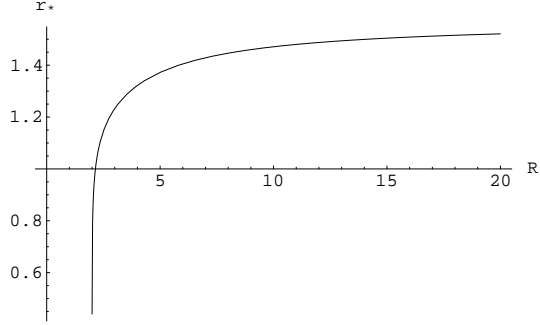}}
\caption{Tortoise coordinate $r_*$ position of the shell 
as a function of $R$. 
Note that the shell moves to $r_*=-\infty$ as $R\rightarrow r_H=2$. }
\end{figure}

\subsection{Quantization}

Given the limitations of the stress tensor as a diagnostic of
properties of a collapsing configuration, we next turn to the calculation
of the two-point function.  This proceeds via canonical quantization of the
field $\phi$.  In particular, $\phi$ can in general be expanded in a
general orthonormal basis
of modes $\left\{\phi_\alpha\right\}$ as
\be
\phi = \sum_\alpha\left( \phi_\alpha a_\alpha + \phi_\alpha^* a_\alpha^\dagger \right)
\label{phiexp}
\en
where $a_\alpha$, $a_\alpha^\dagger$ are the corresponding annihilation and
creation operators.  Definition of such a basis and corresponding ladder
operators also serves to define a vacuum,
\be
a_\alpha |0\rangle  =0\ .
\en
The two-point function is then given by
\be
\langle \phi(x)\phi(y) \rangle = \langle 0| T\left[ \phi(x)\phi(y) \right]
|0\rangle \ .\label{twopoint}
\en

Choosing the basis $\left\{\phi_\alpha\right\} = \left\{ \varphi_{n
l\vecm} e^{-i\omega t} Y_{l\vecm}/\sqrt{2\omega_{nl}} \right\}$ 
given in the last subsection
therefore defines a vacuum
$|0\rangle_{\rm shell}$ and
two-point function for the shell configuration.  The latter takes the form 
\beq
\langle \phi(x)\phi(x')\rangle_{\rm shell}&=&{1\over i}
\sum_{n l \vec{m}} Y_{l{\vec m}}({\hat e})
Y^*_{l{\vec m}}({\hat e}')\int \fr{d \omega}{2 \pi} e^{-i\omega (t-t')} 
\fr{\varphi_{n l \vec{m}}(r) \varphi^*_{n l \vec{m}}(r')}
{\omega_{nl}^2-\omega^2-i \epsilon}\nonumber\\
&=&\theta(t-t')
s(x,x')+\theta(t'-t)s^*(x,x')\label{twosh}
\enq
where ${\hat e}$ and ${\hat e}'$ are unit vectors giving the directions of
$\vec x$ and ${\vec x}'$, and 
\be
s(x,x')=\sum_{n l \vec{m}} Y_{l{\vec m}}({\hat e})
Y^*_{l{\vec m}}({\hat e}') {e^{-i \omega_{nl}(t-t')} \over 2\omega_{nl}}
\varphi_{n l {\vec m}}(r) \varphi_{n l {\vec m}}^{*}(r') \ .
\en

One way of characterizing a vacuum is in terms of local observations
performed, for example, by an observer carrying an Unruh
detector \cite{Unruh:1976db}.  Recall that an
Unruh detector can be thought of as a simple quantum-mechanical system,
e.g. a harmonic oscillator
carried by a
local observer.  The harmonic oscillator lagrangian contains a coupling to
the field of the form
\be
S_I = \int d\tau Z(\tau) \phi(x(\tau))\ ,
\en
where $Z$ is the oscillator position variable, and $\tau$ is the proper time
along the observer's worldline $x(\tau)$.
In the vacuum $|0\rangle_{shell}$ defined by the shell modes, an Unruh
detector carried by an observer following a trajectory of constant $r$
stays unexcited. 

Indeed, in the limit as the shell approaches the horizon, the state
$|0\rangle_{shell}$ approaches another well known state, the Boulware
vacuum.  With the shell at the horizon, the wave equation (\ref{wave}) is
simply the wave equation in the \sads\ background.  Working in
the coordinates $t,r_*$, we derive the effective potential for
\sads; in particular, Schwarzschild time $t$ defines a notion of
positive frequency, and the resulting Boulware 
modes $\varphi_{\omega
l\vecm}^B e^{-i\omega t} Y_{l\vecm}$ define the Boulware vacuum,
\be
a_{\omega l\vecm}^B |0\rangle_B =0\ .
\en
As in the shell case, observers at constant $r$ do not see particles in
this state.
 
\subsection{Boundary description}

According to the prescription of \cite{Gubser:1998bc,Witten:1998qj}, 
the two-point function of the
boundary operator $\cal O$ corresponding to the field $\phi$ is obtained as
a rescaling of the limit of the two-point function (\ref{twopoint}) as its arguments
go to the boundary.  The rescaling needed follows from the asymptotic form
of the normal modes $\phi_{nl\vecm}$.  In particular, this asymptotic form
depends on the AdS radius but not the black hole mass, and we find
\be
\phi_{nl\vecm}\rightarrow {{k}_{nl}\over \sqrt {2 \omega_{nl}} }
\fr{1}{r^{d}} Y_{l\vecm}
e^{-i\omega_{nl} t} \label{bcorr}
\en
with ${k}_{nl}$ constants given in \cite{Giddings:1999jq}.
Thus the boundary two-point function takes
the form
\beq
\langle \calo(b) \calo(b')\rangle 
\propto& &\lim_{r,r'\rightarrow \infty}(rr')^d
\langle \phi(x)\phi(x')\rangle\nonumber\\ 
& & =
{1\over i} 
\int {d\omega\over 2\pi}\sum_{nl\vecm} e^{-i\omega(t-t')} 
k_{nl}^2 {Y^*_{l\vecm}(\ehat) Y_{l\vecm}(\ehat')\over \omega_{nl}^2
-\omega^2 -i\epsilon}\ .\label{bdytwo}
\enq

The spectrum of excitations on the boundary can of course be read off from
the frequencies $\omega_{nl}$.  As the shell
reaches the horizon, these become continuous, corresponding to a continuum
in the boundary theory.  Similar behavior was found in
\cite{Giddings:1999zu} in the case of a shell of D3 branes collapsing into
a black brane.

However, note that this shell vacuum is very different from the physical
state we expect to form outside a black hole, which is described by the
Hartle-Hawking state.  In the case of extremal configuration of D3 branes,
there is no corresponding issue -- the Hawking temperature is zero, and the
Boulware and HH states correspond.  However, in the more physical
non-extremal 
case of
D3 branes with some initial velocity, as well as in the present case, the
states are different.  We would like to better understand the transition
from one to another, and the corresponding boundary physics of thermalization.
We begin by describing the HH state in more detail.

\section{The Hartle-Hawking Green function and thermalization of the shell
state}

\subsection{The Hartle-Hawking state and Green function for 
\sads}

We begin by describing properties of the Hartle-Hawking (HH) 
state and Green
function for
\sads.  Recall that there are several equivalent perspectives on
the HH state and Green function.

In the first perspective, we chose 
a specific set of modes in the
expansion of the field, (\ref{phiexp}), and define the vacuum to be
annihilated by the corresponding annihilation operators.
These modes are taken to have positive frequency with
respect to an affine parameter along the horizon, or equivalently in
the time defined in terms of Kruskal coordinates, that is, they are positive
frequency as seen by a
freely falling observer crossing the horizon.  

In the case of Schwarzschild
in a Minkowski background, one must in fact choose two sets of modes to
form a basis.  The
first set can be defined to form a basis for solutions to the wave equation
that 
vanish at past null infinity $\scrim$ and are   
purely positive frequency with respect to Kruskal time 
on the past horizon $\hmi$.  The second set is a basis for solutions that
vanish on $\hmi$ and are positive frequency with respect to Schwarzschild
time at $\scrim$.  Together these two sets of incoming modes
form a basis for all
solutions and define a vacuum.  
One may alternately define a basis of outgoing modes
that is positive frequency
at the future horizon $\hpl$ and future null infinity $\scrip$, and a
corresponding vacuum.  

In \sads, if we restrict to normalizable solutions, the 
boundary
conditions 
mean that in the far past complete
Cauchy data is specified on $\hmi$, or alternately 
in the far future it is specified
on $\hpl$.  In this case, we can choose a complete set of positive
frequency modes on the past horizon, and define a corresponding vacuum 
$|0\rangle_-$.  Alternately we may choose a basis that is positive frequency
on the future horizon, and define a corresponding vacuum $|0\rangle_+$. The
different definitions of the vacuum suggest an ambiguity in specifying the
vacuum, and in defining the HH Green function.  However, in 
\cite{Gibbons:1978pt} Gibbons and Perry use an analyticity argument to show
that these are in fact the same vacuum for the Schwarzschild solution in a
box; in the present case,
AdS supplies a gravitational definition of a box.  Thus $|0\rangle_+=
|0\rangle_-\equiv|0\rangle_{HH}$, and the HH Green function is
\be
G_{\rm HH}(x,x') = {}_{\rm HH}\langle 0| 
T\left[\phi(x)\phi(x')\right] |0\rangle_{\rm HH}\ .
\en

A second definition of the HH state and Green function follows
the original paper \cite{Hartle:1976tp} more directly.  Continuing the 
complete \sads\ metric (\ref{adsc}) to euclidean signature gives
the metric
\be
ds^2=f_{+}d\tau^2+dr^2/f_{+}+r^2d\Omega^2_{d-1} \label{eucsch}
\en
with topology $R^2 \times S^{d-1}$.  
Slicing this metric along the slice $\cals$  of $\tau =0$,
$\tau=\pi$ (see fig.~6) gives a spatial metric that agrees with spatial slices of
constant Schwarzschild time in the full Kruskal extension of lorentzian
\sads.  A Green function may be defined directly on the
euclidean manifold (\ref{eucsch}) and then its arguments may be
analytically continued to points in lorentzian \sads.  The
result is again the HH Green function; equivalence follows from the
analyticity properties of the continued lorentzian HH Green function, as
discussed in \cite{Gibbons:1978pt}.
\begin{figure}
\centering
\scalebox{1}{\includegraphics{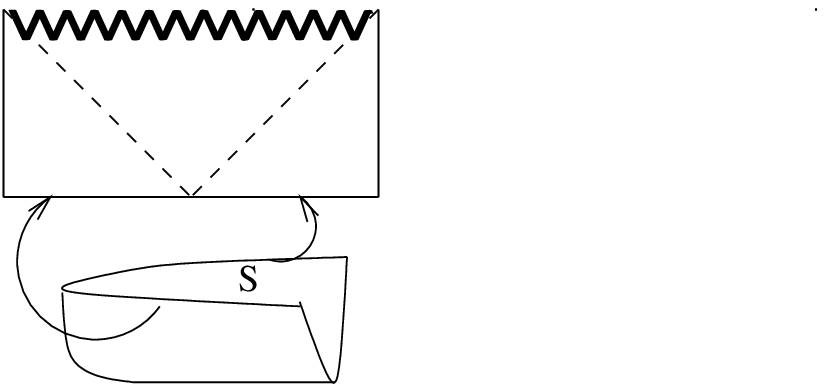}}
\caption{Matching of Euclidean and lorentzian \sads. 
The corresponding spaces are identified along the surface ${\cal S}$. }
\end{figure}

A third, related, perspective sees the HH state and Green function as
thermal with respect to observers traveling along trajectories of
constant Schwarzschild radius $r$, whose proper time is $\propto t$.  This
thermal character is particularly simply elucidated in 
\cite{Jacobson:1994fp,Jacobson:1995ak}, following earlier work of 
Israel\cite{Israel:1976ur}
and Unruh\cite{Unruh:1976db}.  In particular, the euclidean functional
integral defines the HH state on the slice $\cals$.  A Schwarzschild
observer sees only half of this slice; label the states seen by this
observer as $|\phi_R\rangle $.  To specify a state on the full slice
$\cals$, one must also specify the state $|\phi_L\rangle $ as seen by a Schwarzschild
observer in the left quadrant, and so one can think of states on $\cals$ as
superpositions of states of the form $|\phi_L\phi_R\rangle $ in the tensor
product Hilbert space.  The euclidean functional integral defining the HH state can
be written in terms of the Schwarzschild hamiltonian $H_S$
which 
generates evolution
in the angular direction in the euclidean geometry, and so in the $L\times
R$ basis the HH state takes the form
\be
\langle \phi_L\phi_R|0\rangle_\hh \propto \langle \phi_L | e^{-\pi H_S}
|\phi_R\rangle \ .
\en
Since the HH Green function is the expectation value of operators defined
in the right quadrant, this expectation value reduces to a trace over the
left modes.  This yields the thermal form for the Green function,
\be
{}_\hh\langle 0| T\left[\phi(x) \phi(x')\right]|0\rangle_\hh \propto Tr
\left( e^{-2\pi H_S} T\left[\phi(x) \phi(x')\right]\right)\ .
\en

Specifically, work in a basis for Schwarzschild modes of the form given in 
eq.~(\ref{schmode}).  In this basis the HH Green function then takes a
form originally given in \cite{Candelas:1980zt}
\be
G_\hh(x,x')=\theta(t-t')s_\hh(x,y)+\theta(t'-t)s_\hh^*(x,y) \label{hhgf}
\en
with
\beq
s_\hh(x,y)&=&\sum_{l \vec{m}} \int_0^\infty    {d\omega\over 2\omega} 
\Biggl[\fr{e^{-i \omega(t-t')} 
\varphi_{\omega l \vecm}(r)Y_{l\vecm}({\hat e}) 
\varphi_{\omega l \vecm}^{*}(r')Y_{l\vecm}^*({\hat e}') }{1-e^{-\beta
\omega}}\nonumber\\&+& \fr{e^{ i \omega(t-t')} 
\varphi_{\omega l \vecm}^*(r)Y_{l\vecm}^*({\hat e}) 
\varphi_{\omega l \vecm}(r')Y_{l\vecm}({\hat e}') }{e^{\beta
\omega}-1}\Biggr]
\enq
and
\be
\beta=\fr{4\pi r_Hb^2}{d r_H^2+(d-2)b^2}\ .
\en
The corresponding boundary Green function is derived as in (\ref{bdytwo})
and takes the analogous form
\be
\langle \calo(b) \calo(b')\rangle_\hh \propto 
\theta(t-t')s_\hh(b,b')+\theta(t'-t)s_\hh^*(b,b') 
\en
with
\be
s_\hh(b,b')=\sum_{l\vecm} \int {d\omega\over 2\omega} k_{\omega l}^2 \left[ 
{Y_{l\vecm}(\ehat) Y_{l\vecm}^*(\ehat') e^{-i\omega(t-t')}\over
1-e^{-\beta \omega}}
+ {Y_{l\vecm}^*(\ehat) Y_{l\vecm}(\ehat') e^{i\omega(t-t')}\over
e^{\beta \omega}-1}\right]
\en
where $ k_{\omega l}$ are the continuum analogs of the constants defined in
(\ref{bcorr}). 

These Green functions clearly differ,
by virtue of being thermal, from (\ref{twosh}), (\ref{bdytwo}) 
in the limit where the shell
approaches the horizon.

\subsection{Thermalization and approach to the Hartle Hawking state}

The relationship between the shell/Boulware states that we have previously
described and the Hartle-Hawking state that one expects to observe once a
black hole has formed appear to be close to the heart of the black hole
information paradox.  In particular, the shell/Boulware state can be
thought of as a pure state.  We can imagine physically constructing it by
quasistatically 
lowering steel plates close to the black hole horizon and bolting them
together.  However, if we sever the bolts, the shell collapses to a black
hole which has as its external description the Hartle-Hawking state.  This
transition from an obviously pure state to an apparently thermal (but still
presumably pure) 
state is expected to  have a boundary description as a transition between a
dynamical pure state and an apparent thermal state in the gauge theory.  If
we could explicitly describe this transition we would know where the
information of the black hole is hidden, and may be able to better
understand how it is revealed when the black hole evaporates.  

It should also be noted that there is a direct relationship between the
shell/Boulware state and the Hartle-Hawking state.  In particular,
Mukohyama and Israel\cite{Mukohyama:1998rf} have argued that if one starts
with a state of the shell/Boulware form and then adds on top of it thermal
radiation to give what they call a ``topped-up Boulware'' state, this has
the same properties (e.g. stress tensor, correlators) as the Hartle-Hawking
state.  In fact, this relationship is evident from (\ref{hhgf}) which
exhibits the HH Green function as a thermal Green function in the basis
appropriate to the Schwarzschild observer.
The important
question is how the state makes the transition from the shell/Boulware
state to the Hartle-Hawking state.  It is tempting to conjecture that this
is by radiation from a collapsing 
shell, but we have not yet found a direct description of this
process.

\section{Discussion}

The dynamical shell problem remains difficult.  The shell/Boulware state
and HH state are very different, and it is not obvious how to physically
connect them.  The problem of studying the dynamical shell is akin to a
moving mirror problem \cite{Davies:hi},  
and perhaps further insight can be gained from
studies of that problem and/or from more recent treatments of Hawking
radiation as a dynamical process\cite{Melnikov:2001ex}.
Even in principle it is difficult to identify the
degrees of freedom where the black hole entropy is encoded.

A closely analogous problem was outlined in the introduction:  one
may study collapse of a spherical shell of D3 branes, as in 
\cite{Giddings:1999zu}, to form a
black brane.  Here we have a clearer view of the degrees of freedom; the
positions of the D3 branes correspond to scalar vevs in the field theory,
and oscillations of the branes likewise translate into boundary fields.
The case of D3 branes presents some slight modifications to the above
story.  The collapse of a dynamical shell of $N$ D3 branes likewise reveals
the emergence of the continuum of frequencies and here we also  
expect to find apparently thermal final state analogous to that of
Hartle and Hawking, with temperature
\be
T_H=\left({8\over 3\pi^2 N^2} \mu\right)^{1/4}
\en
where  $\mu$ is the mass  density
above 
extremality.
The question is how to understand the transition from an initially pure
state to this thermal state.  In the case of D3 branes, there are a few
more clues about the relevant degrees of freedom.  In particular, one can
excite string states on the D3 branes, 
and think of  the apparent entropy of the black brane as entanglement entropy
between the degrees of freedom outside the branes and the excited string
states on the branes.  This picture nicely corresponds with computations of
black brane entropy\cite{Gubser:1996de} via counting of string states on the D3 branes.
On the gauge theory side, it would be very interesting to understand this
process in more detail.  The starting configuration is a point on the
Coulomb branch with an initial velocity for the adjoint scalar vevs.  As
the vevs reach the origin, the system should undergo thermalization of the
gauge degrees of freedom.  This should be the 
dual description of the transition to the radiating Hartle-Hawking state.
But then an even more puzzling question
is how to map this description onto a description appropriate
to an observer falling into the black brane.  In particular, it  is very difficult to
understand what variables in the apparently thermal boundary state
could describe the large coherent internal region
accessible to the infalling observer.

Returning to the case of a shell collapsing to form a black hole, if the
initial black hole is small enough, it will subsequently Hawking radiate
back to the geometry of global AdS, with some apparently 
thermal radiation.  According to the AdS/CFT correspondence, the dual
boundary process is described by unitary evolution, and so if the
correspondence really holds at this level of detail, the final bulk state should
also be a pure state.  It would be very interesting to go further and
understand the boundary description of the evolution of the black hole to
zero mass, and to explain where the hidden correlations that make the
apparently thermal state pure lie in the final state.  If it is possible to
track the information flow in this fashion, and in particular to find the
explicit corrections to Hawking's original derivation\cite{Hawking:sw},
that would finally provide an
understanding of how string/M theory resolves the black hole information
paradox.  One can also ask a parallel question for the D3 brane shell.  If
one begins with nonzero initial velocity, then a non-extremal black brane
forms, and then will Hawking evaporate back to an  extremal brane,
corresponding to AdS in the Poincare parameterization.  In this process,
since the entropy of the brane decreases, information should be present in
the Hawking radiation that has been emitted.  One would very much like to
see evidence for this information in the dual boundary description of the
process of collapse followed by evaporation of the brane.

Note added:  after this paper appeared, 
refs.~\cite{Ilha:1996tc,Lemos:1998iy,Ilha:1999yn} were brought to our
attention; these discuss gravitational collapse in with non-zero
cosmological constant,
and in particular collapsing shells and Oppenheimer-Sneider
solutions.

\section*{Acknowledgements}

The authors wish to thank J. Hartle, 
G. Horowitz, P. Kraus, S. Ross, and J. Traschen
for useful
conversations. This work was supported in part by the Department of Energy
under Contract DE-FG-03-91ER40618,  by the 
National Science Foundation under Grant No. PHY99-07949, and by a Graduate
Division Dissertation Fellowship from UCSB. 

\renewcommand{\theequation}{A-\arabic{equation}}
\setcounter{equation}{0} 
\section*{ Appendix: The  anti-de Sitter Oppenheimer-Snyder solution}

In this appendix we present another example of gravitational collapse in
AdS: that of a ball of pressureless dust. 
In $d+1$ dimensions, this solution has 
energy-momentum tensor 
\be
T_{00}=\rho_0/a(\tau)^d\ ,\ T_{0i}=T_{ij}=0
\en
where $\rho_0$ is a constant related to the mass of the ball of dust.
The collapsing dust creates a black hole as shown in fig.~7. 
\begin{figure}
\centering
\scalebox{0.4}{\includegraphics{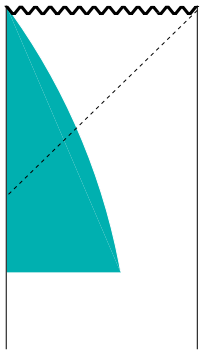}}
\caption{Collapsing ball of dust in anti de-Sitter space. }
\end{figure}
The internal  metric takes Friedmann's form
\beq
ds^2=-d\tau^2+a(\tau)^2(d\chi^2+A(\chi)^2d\Omega_{d-1}^2)
\label{freid}
\enq
with
\be
A(\chi)=\sin(\sqrt{k} \chi)/\sqrt{k},
\en
and $k=-1,0,1$ corresponds to open, flat and closed
spatial geometries, respectively.

Due to the spherical symmetry, the Einstein equation for the 
metric (\ref{freid}) reduces to a single ordinary differential equation:
\be 
{\dot a}^2 + k + {a^2\over b^2} = {2\kappa\rho_0\over d(d-1) a^{d-2}}\
.\label{en_ball} 
\en
This is solvable by quadrature,
\be
\tau = \int_{a_0}^a da \left[{2\kappa\rho_0\over d(d-1) a^{d-2}}-k-{a^2\over
b^2}\right]^{-1/2} 
\en
where $a_0$ is an integration constant.

Outside the ball of dust, the external metric is the solution of the 
vacuum   equation, which by Birkhoff's theorem is $d+1$ dimensional 
\sads:
\be
ds^2=-f dt^2+dr^2/f+r^2 d\Omega_{d-1}^2 \label{mmmet}
\en
with
\be
f=1-m/r^{d-2}+r^2/b^2\ .
\en

These two solutions should be matched on the world-tube of the surface of
the ball.  In the internal coordinates, this is given by $\chi=\chi_0$, and
in the external coordinates by a trajectory $R(t)$.  We must match the
induced metric and extrinsic curvature\cite{Israel:1966rt}.  The metric of
the world-tube is
\be
d\sigma^2 = -d\tau^2 + a^2(\tau) A^2(\chi_0) d\Omega_{d-1}^2
\en
in the internal geometry, and
\be
d\sigma^2 = -fdt^2 + \left({dR\over dt}\right)^2 {dt^2\over f} 
+ R^2 d\Omega_{d-1}^2
\en
in the external geometry.  Matching these metrics then yields
\be
R(t)= a(\tau) \sin(\sqrt{k} \chi_0)/\sqrt{k}\label{Mone}
\en
and
\be
{d\tau\over dt} = \sqrt{f^2(R)-\left({dR\over dt}\right)^2   
\over f}\ .\label{Mtwo}
\en
The extrinsic curvature, which is the 
symmetrized 
covariant derivative of the unit normal to the surface,
\be
K_{\mu \nu}=-\fr{1}{2}(\nabla_\mu n_\nu+\nabla_\nu n_\mu)\ ,
\en
can also be computed in the internal and external geometries.  In the
internal geometry, we find
\beq
K_{\tau \tau}=0 \\
K_{\theta \theta}=-a(\tau)A(\chi_0)\fr{dA}{d\chi}\ .
\enq
In the external metric (\ref{mmmet}), the independent components of the 
extrinsic curvature are
\beq
K_{\tau \tau}=\frac{\ddot{R}+f'(R)/2}{\sqrt{f+\dot{R}^2}}\\
K_{\theta \theta}=-R\sqrt{\dot{R}^2+f}\ .
\enq
where dot denotes derivative with respect to $\tau$.
To the conditions (\ref{Mone}), (\ref{Mtwo}), curvature matching adds
\beq
\cos(\sqrt{k}\chi_0)=\sqrt{\dot{R}^2+f(R)}\ .
\label{match2}
\enq
Comparing  (\ref{match2}) with the Einstein equation
(\ref{en_ball}), we 
obtain the relation between the mass and the parameters of the dust configuration:
\be
m= {2\kappa\rho_0\over d(d-1)} \left[ {\sin({\sqrt k}\chi_0)\over {\sqrt k}}
\right]^d\ .
\en

\end{document}